\documentstyle[aps,epsfig]{revtex}
\begin{document}
%\textwidth = 120mm
%\textheight 40\baselineskip
%\hoffset=-0.5cm
\voffset=-0.5cm
%%%%%%%%%%%% LOCAL MACROS %%%%%%%%%%%%%%%%%%%%%
\def\bra{\langle}
\def\ket{\rangle}
\def\id{1\kern-.25em\hbox{I}}
\def\C{{I\kern-.60em{C}}}
\def\N{{I\kern-.25em{N}}}
\def\Q{{I\kern-.6em{Q}}}
\def\R{{I\kern-.25em{R}}}
\def\Z{{Z\kern-.5em{Z}}}
\def\cA{{\cal A}}
\def\cD{{\cal D}}
\def\cL{{\cal L}}
\def\cH{{\cal H}}             % Hilbert space
\def\cU{{\cal U}}
\def\cH{{\cal H}}
\def\cQ{{\cal Q}}
\def\cZ{{\cal Z}}
\def\Qp{{~Q_p}}
\def\Zp{\Z_p}
\def\ie{i.~e.\,}
\def\eg{e.~g.\,}
\def\lhs{l.~h.~s.\,}
\def\rhs{r.~h.~s.\,}
\def\semdir{{\mid}\kern-.25em{\times}}
\def\d{\partial}
\def\df#1#2{\frac{\d {#1}}{\d {#2}}}
\def\rot{{\bf \nabla}\times}
\def\lr{{\rm L}^2({\rm R})}
\def\rref#1{(ref{#1})}
\def\hc{\dagger}              % Hermitian conjugation
\def\Ex{{\rm E}}              % Expectation
\def\Cov{{\rm R}}             % Covariance
\def\iff{if and only if}
\newcommand{\be}{\begin{equation}}
\newcommand{\ee}{\end{equation}}
%%%%%%%%%%%%%%%%%%%%%%%%%%%%%%%%%%%%%%%%%%%%%%%%%%%%%%%%%%%%
\title{$p$-Adic physics below and above Planck scales}
\author{M.V.Altaisky\thanks{Also at Joint Institute for Nuclear Research, 
Dubna, 141980, Russia}\thanks{Also at Space Research Institute RAS, 
Profsoyuznaya 84/32, Moscow, 117810, Russia} and B.G.Sidharth\\
B.M.Birla Science Centre, Adarshnagar, Hyderabad
500 063, India}
\maketitle
\begin{abstract}
We present a rewiew and also new possible applications
of $p$-adic numbers to pre-spacetime physics. It is shown that
instead of the extension $\R^n\to\Qp^n$, which is usually implied
in $p$-adic quantum field theory, it is possible to build a model based
on the $\R^n\to\Qp$, where p=n+2 extension and get rid of loop
divergences. It is also shown that the concept of mass naturally
arises in $p$-adic models as inverse transition probability with a
dimensional constant of proportionality. 
\end{abstract}
\section{Introduction}

The results of any physical measurements are always expressed in
terms of {\em rational} numbers: since any measuring equipment works
with finite precision, bounded at least by the quantum uncertainty
principle. The next step --- to the field of {\em real} numbers ---
is a theoretical abstraction valid only at large classical scales.
The geometry of the Planck scales \cite{Wheeler} is, of course, unknown
but there is a strong theoretical
bias that the geometry of these scales should be non-Archimedian.
Indeed, who and how could compare two distances then?

It has now come to be realised that the smooth spacetime continuum that is
used even in quantum field theory is an approximation
\cite{Rovelli91,Kreinovich}. Space-time at the small scales appears to be
granular. Indeed the very
concept of the usual space and time in quantum theory is questionable
\cite{Sonego,Parker}. The fact that time itself would have a minimum
irreduceable unit, the chronon, was proposed quite sometime back
\cite{Schild,Caldirola}. These ideas find a
culmination in the recent model of the quantum-mechanical black holes
(QMBH) \cite{qmbh1,qmbh2}. In this model
the Compton wavelength and the corresponding time interval which is the same
as the minimum time interval, the chronon refered to arise as the natural space
time cutoffs -- the Planck scales being a special case thereof. It is also
worth noting that as pointed out by Wheeler \cite{Wheeler} and others, it is a
misconception to think that space and time are on the same footing. Rather
our perception and hence description of the universe is one of "all space"
at "one instant", that is effectively time is a parameter. Space and time
are on the same footing in a stationary scenario, that is for a closed system
as a whole \cite{Page,Sidharth}. This is brought out for example
in the fluctuating foam like structure of space time at Planck scales
\cite{Wheeler}.

Besides,there is a more rigouros (or may be more metaphysical) requirement
that physical laws should be expressed in a coordinate-free
language, \ie as relations between some objects and sets
(See \eg\cite{Fin95} for a comprehensive review).
This geometry  --- the quantum topology --- is unknown as yet,
and it is more philosophical than a physical question
whether or not it can be discovered in detail.
However it seems quite reasonable that it should be
non-Archimedian, for there is no way to compare the distances at
Planck scales.
There are different models leading to non-Archimedian
geometry. To some extent, we may say that the basic idea is
that any set of objects can be labeled by an infinite
series of integers. In $p$-adic
models it is believed that $p$-adic numbers play an important
role at the fundamental scales.

From the first glimpse it may seem that a $p$-adic number
$$
x=\sum_{k=k_{min}}^\infty a_k p^k,\quad 0\le a_k<p,$$
(where $p$ is prime) is just another representation  of
an infinite sequence $\{ a_k \}_{k_{min}}^\infty$. In fact,
is there any difference whether $p$ is a prime or just an integer?
Physical intuition suggests no difference here. However, the
mathematical truth is that only prime integers label
different topologies available at continuum\cite{BF}.
This fact seems to be of great physical importance. One may play
with mathematics and methaphysics of the Planck scales as long
as possible, but the results of all physical observations  are obtained
at the energies much below the Planck ones, therefore the correspondence
between the existing and well tested theories (such as quantum
electrodynamics) and that from the waiting list of quantum gravity
can be checked for correspondence only in a continuous limit. (This
however does not prohibit pure theoretical studies at Planck scale
energies, which may lead to some cosmological consequences.)

The paper is organized as follows. In {\em section 2} we recapitulate
the basic facts about $p$-adic numbers. In {\em section 3} we give a
critical review of quantum models constructed on the $D$-dimensional
$p$-adic space $\Qp^D$.
In {\em section 4} we present our geometric
approach to $p$-adic quantum field theory and its possible links to the
space or space-time metric. Some possible consequence of
all these things are presented in {\em last section}.

\section{Basic facts about $p$-adic numbers}
The results of any measurements, as was already mentioned,
can be expressed in terms of rational numbers. The construction of a
theoretical model, first of all a description in terms of differential
equations, ultimately requires an extension of this field. The first
extension
(made by Dedekind, viz. equivalence classes) means the incorporation of
rational numbers. It is a completion of the field $\Q$ with respect to
the standard norm $\Q \stackrel{|\cdot|}{\rightarrow} \R$. This completion
however does not exhaust all possibilities. It is also possible to extend
the field $\Q$ using the {\it p-adic norm} $|\cdot |_p $
(to be explained below):
\be
\R \stackrel{|\cdot|}{\leftarrow} \Q
\stackrel{|\cdot|_p}{\rightarrow} \Qp.   \label{ost}
\ee
This completion is called the field of $p$-adic numbers $\Qp$.
No other extensions of $\Q$ except these two exist due to the
Ostrovski theorem \cite{Ost}.

The $p$-adic norm $|\cdot |_p$ is defined as follows.
Any nonzero rational number $x \in \Q$
can be uniquely written in the form
\be
x = {m\over n}p^\gamma,  \label{pdec}
\ee
where integers $m$ and $n$ are not divisible by the prime integer $p\ne1$,
and $\gamma$ is an integer.  The decomposition (\ref{pdec}) provides
a possibility to supply the field $\Q$ with the
norm
\be
|x|_p = {\left| {m\over n}p^\gamma \right|}_p \label{pnorm}
\stackrel{def}{=} p^{-\gamma}, \qquad |0|_p \stackrel{def}{=} 0,
\ee
different from the standard one. The algebraic closure of the field
$\Q$ in the norm $|\cdot|_p$ forms the field of $p$-adic numbers
$\Qp$.

Any $p$-adic number can be uniquely written in the form
\be
x = \sum_{n=k_{min}}^{\infty} a_n p^n,
\qquad a_n \in \{0,1,\ldots,p-1\}, \quad k_{min} > -\infty.
\ee
It is easy to check that
$|xy|_p = |x|_p|y|_p,$ but $|\cdot|_p$ is stronger than $|\cdot|$:
\be
|x+y|_p \le max (|x|_p,|y|_p) \le |x|_p+|y|_p
\ee
and induces a non-Archimedian metric
\begin{eqnarray}
d(x,y) &:=& |x-y|_P \label{pdist}\\
\nonumber d(x,z) &\le& max (d(x,y),d(y,z)) \le d(x,y) + d(y,z),
\end{eqnarray}
often called an {\em ultrametric} \cite{RV86}.
With respect to the metric (\ref{pdist}) the $\Qp$ becomes a
complete metric space. The maximal compact subring of $\Qp$
\be
\Z_p = \{ x \in \Qp : |x|_p \le 1 \}
\ee
is referred to as {\em a set of p-adic integers}.
The field $\Qp$ admits a positive Haar measure,
unique up to normalization
\be
d(x+a) = dx, \quad
d(cx)  = |c|_p dx, \quad x,a,c \in \Qp.
\label{hmp}
\ee
The normalisation is often chosen as
$\int_{\cZ_p} dx \equiv 1$.

The geometry induced by the distance $|x-y|_p$ is quite different
from the Euclidean one: all $p$-adic triangles are equilateral;
two $p$-adic balls may either be one within another or disjoint.

There is no unique definition of differentiation in the field $\Qp$,
but the Fourier transform exists and is used in $p$-adic field
theory to construct the (pseudo-)differential operator
$$\nabla \phi(x) \to |k|_p \tilde \phi(k).$$
The construction of the $p$-adic Fourier transform is essentially
based on the group structure of the field $\Qp$, viz. the group
of additive characters
$$ \chi_p(x) := \exp \bigl( 2\pi\imath \{ x \}_p \bigr),
\quad \chi_p(a+b) = \chi_p(a)  \chi_p(b),$$
(where $\{ x \}_p$ denotes the {\it fractional part} of $x$:
$\{ x \}_p = a_{min} p^{k_{min}} + \ldots + a_{-1} p^{-1}$),
is used to construct the Fourier transform 
\be
\tilde \psi(\xi) = \int_{\Qp} \psi(x) \chi_p(\xi x)dx,\quad
\psi(x) = \int_{\Qp} \tilde \psi(\xi) \chi_p(-\xi x)d\xi.
\ee
The $n$-dimensional generalization is straightforward
%\begin{eqnarray*}
$$
\Qp\to\Qp^n, \quad
x\to(x_1,\ldots,x_n), \quad
\xi\to(\xi_1,\ldots,\xi_n), \quad
\xi x\to(\xi,x)=\sum_i\xi_i x_i.
$$
%\end{eqnarray*}

\section{$p$-Adic quantum field theory}
As any field theory starts from the action functional
$S[\phi,\d\phi,\ldots]$, so does the $p$-adic theory. At the first
glance, it is just a bare substitution of the numeric field
$\R^D\to\Qp^D$, and hence for the generation functional
$Z[J]\to Z[J]|_{\Qp^D}$:
\be
\begin{array}{lcl}
Z[J]|_{\Qp^D} &=& \int \cD\phi \exp \Bigl(\int_{\Qp^D} J\phi d^Dx + S[\phi] \Bigr) \\
\label{gf}
S[\phi] &=& \int_{\Qp^D} \cL (\phi,\d\phi,\ldots) d^Dx,
\end{array}
\ee
where all integrations in (\ref{gf}) are taken in the Fourier
space, as in usual quantum field theory. Referring the reader
to \cite{smr,padic} for the detailed account of the
Green functions and the Feynman expansion calculation, we have only
to note that the theory (\ref{gf}) inherits (of course in
milder form) the divergences of the loop integrals.
The integral $$\int_{\Q_p^D} \frac{d^Dk}{|k^2|_p+m^2}$$
is divergent for $D\ge2$.
%(For the sake of simplicity a theory of a scalar field $\phi(x)$
%with a polynomial interaction is considered.)

Up to now we did not touch upon the question {\it why} it may be possible to
substitute a numeric field in a field theory (scalar for simplicity)
defined over a (pseudo-)Euclidean space by means of a certain action
functional. First, it can be easily noticed that the availability of
the extension $\Q \to \Qp$, even taken together with
the principles of quantum field theory, does not suggest any particular
value of $p$. Fortunately, as it follows from number theory,
{if all prime bases are taken togeter} their collection resembles
the field of real numbers, viz.
$$
\prod_{p\in prime} \chi_p(\omega t-kx)=\exp(2\pi\imath(\omega t - kx))
$$
and in this sense the {\it real free particle is a product of an infinite
number of p-adic plane waves}. The other but similar argument, which
has caused a deep interest in $p$-adics, is that the formulae for
the string amplitude have the same form as $p$-adic adelic products.

\section{Geometric approach to $p$-adic quantum field theory}
The question, from what and how our Universe
and hence spacetime have emerged is a deep one. The standard tool of modern
physics applied to this problem is a retrospective analysis. Starting
from todays structure of the Universe we extrapolate its geometrical
properties back in time using the results both general relativity and particle
physics. The extrapolation stops at  the Planck scales, where the
problem whether or not we can go further down with our metric theories
stops our consideration. However, we can turn the problem upside down.
If the Universe {\it has emerged}, it should emerged from One
{\it something}. At the same time a large variety of different objects
exist now.
So, there is a problem of how many could emerge from one. It is not only a
philosophical, but also a physical problem.

If we reject the idea of {\it a priori} background space, then
there is no geometrical space when there is only one object, and
there is no time without a changing variety of material objects.
So the {\it appearence
of space and time requires certain relations between objects}. We may think
at least of two kinds of such relations: (i) the relations between objects
belong to a certain set, and (ii) between "parents" and "descendants".
The latter type of relations provides the multiplication of the initial
set. At quantum level the "parent" does not exist any longer after giving
ancestors, \eg $\gamma \to e^+e^-$.

To start with our geometrical approach, let us consider a
two dimensional sphere $S^2$ and find out what other coordinates
except for well known ones ($\theta,\phi$) can be used to label
the points on the sphere. To do this, we use the fact that an
$n$-dimensional sphere can be considered as a boundary of an
($n+1$)-dimensional simplex. For two dimensions this implies that
we can just take a 2D simplex --- a triangle, --- identify its
vertices ($A=B=C$, see Fig.1), glue edges, and then getting a 3D
simplex put it onto the sphere $S^2$.
\vskip1mm
\begin{center}
\begin{figure}[ht]
\vskip2mm
{\centering \epsfig{file=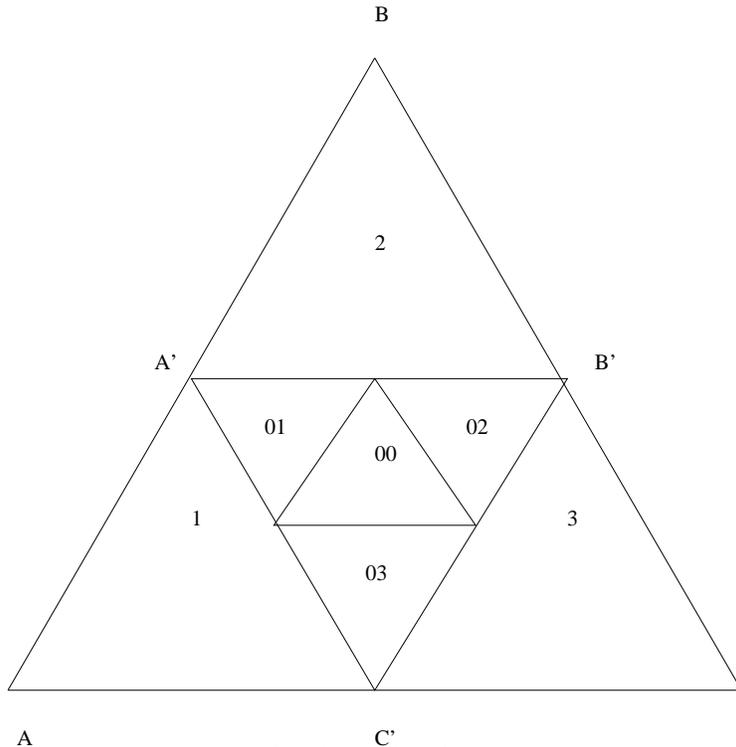, width=10cm} }
\caption{The partition of a d-dimensional simplex (d=2) into (d+2) equal
parts. Identification A=B=C and gluing edges are performed to put it over 
the sphere $S^2$.}
\label{part:pic}
\end{figure}
\end{center}
We can divide our simplex
into p=4 equal equilateral triangles and label them by 0,1,2,3.
The initial triangle is therefore subdivided into $p=d+2$ equal
parts; since the same procedure can be also applied to its parts
we can move down ad infinum and label these parts as ($a_0,a_1,\ldots$,
where $0\le a_i < p$). If we ascribe a measure $\mu(T_0)=1$ to the
initial triangle, when at first stage we have $p$ triangles of
the measure $1/p$ eacn, at the second stage $p^2$ triangles of the measure
$1/p^2$ and so on. To meet this natural definition of the measure, we can
label the sequences $\{ a_i \}$ by $p$-adic fractional numbers
$x\in \Zp^{-1}$.

Then the measure of the (0,1) triangle for instance,
(see Fig.1) labeled by
$x=1p^{-2} + 0p^{-1}$, will be $\mu(0,1) = |x|_p^{-1} = p^{-2}$.
One may also label the multiplicative partition process by $p$-adic
integers $x\in \Zp$. In this case the measure can be taken as
$\mu(x) = |x|_p$. In any case it is possible to locate any point of
the sphere by an infinite sequence of its triangle vicinities of
diminishing measure, which are in one-to-one correspordence to the
space of $p$-adic integers $\Zp$.

Here we would like to touch upon the question of the physical
origin of that we call a {\it continuous manifold}. If there
are only a few objects (with some relations between them) we cannot
speak of a manifold. For example, the partition of
a triangle seen above could be to some extent identified with 3 quarks (and
something inbetween them) which constitute a nucleon, but nothing can
be asserted about a continuous space here. If a collection of many more
objects is considered and there is a dense web of relations between them,
then we can start a continuous approximation -- a manifold.

If we believe that the basic principles of quantum field theory
developed for Euclidean space are also valid for such nets we
can apply the Green functions, loop integrals etc., but there is no
general need to project a $D$-dimensional Euclidean space or a
$D$ dimensional sphere onto this network (to that labeled by
$p$-adic numbers, in particular). All matrix elements, the Green functions
etc. can be evaluated on the $\Qp$ which label the triangle partitioning
of the two-dimensional sphere $S^2$, the same to some extent can
be done by the extensions $S^D \to \Qp,p={D+2}$ for higher dimensions.
As an example let us consider the loop integral
\be
\int_{\Zp} \frac{dk}{|k|_p^2 + m^2} =
\sum_{\gamma=-\infty}^0 \int_{S_\gamma} \frac{dk}{|k|_p^2 + m^2} =
\left({1-{1\over p}}\right) \sum_{\gamma=-\infty}^0 p^\gamma
\frac{1}{p^{2\gamma}+m^2}
\label{i}
\ee
The inegral is evidently finite for any $p$. So we have just demonstrated
the possibility of getting rid of loop divergences in $p$-adic field theory.

It is worth noting at this point that the self interaction which leads
to infinities both in the classical theory of the electron and in QED, is
infact perfectly meaningful and gives rise to the mass in the theory of
the QMBH refered to earlier \cite{Sidharth}, as we will see later.

Now let us give a geometrical interpretation to the integration formula
(\ref{i}). The sum of its right side it is due to the isotropy of the
integrand $f(k) =\frac{1}{|k|_p^2 + m^2}$ which depends on the $|k|_p$
only, but not on the $k$ itself: $f(k)$ is a constant on $p$-adic
circles $S_\gamma = \{ x\in \Qp | |x|_p = p^\gamma \}$. Since each
of these circles contains exactly $(p-1)$ points, the sum can be easily
evaluated. This is an analogue of the integration of the spherically
symmetrical functions in Euclidean space $d^3r \to 4\pi r^2 dr$.

The case of the $p$-adic partitioning of the sphere may first seem
very specific. More generaly, any point in a physical space can
be located by the nested set of its vicinities:
\be
\ldots \subset V_1 \subset V_0.
\label{nest}
\ee
The difference between two nested vicinities (balls) can be referred to as
a circle $S_\gamma = V_{\gamma-1}-V_\gamma,$ exactly as in the $p$-adic case.
The spaces $V_i$ of the nested family (\ref{nest}) do not form a
$\sigma$-algebra, their differences $S_i$ do, and that is why the integration
is defined on $\sum_i S_i$.

Up to now we dealt with mathematical constructions only. The question
is are there any physical counterparts? Evidently
the description of the Universe in terms of locally Euclidean coordinates
is not unique. The location of any physical object in the
Universe can be also defined by the nested set of vicinities (\ref{nest}):
viz. any quark is inside a nucleon/meson, any nucleon inside an atom,
atom inside a galaxy, and so on up to the whole Universe ($V_0$), which
contains everything by definition:
$$ x \in V_m \subset V_{m-1}\subset\ldots \subset V_0.$$
The intersection of each two of these sets should coincide
with one of them: they can not have a partial intersection. The other
property, which is physically required, is that the sets $V_i$, and
hence their differences $S_i$, should have some discrete symmetry. If
all they were structureless, it would be only one quark, one atom, one
galaxy etc. in the Universe.

The situation is much like the structure of $\Qp$, but is not completely
identical to it. There is also the question of how distant galaxies of
our present Universe could be related to a non-Archimedian $p$-adic toy model.
The answer may be as follows. The present state of the Universe is a
result of expansion which has been taken place after the Big Bang. Before
the Big Bang it might have been only a network of relations between some primary
objects emerged from the primary One. Then, due to multiplicative processes
the number of the objects (particles) increased greatly, but some of
the relations between them were inherited and manifest themselves even
in large-scale structures. The distance between different objects, even
between galaxies, may therefore be measured not only by traveling light
waves, but also by the level of their common ancestor in the evolution
process.

It is important to note here, that two simple asumptions: (i) that the
set of vicinities $\{V_i\}$ with which we locate any object is countable
and (ii) that the union
$\cH = \overline{\cup_i V_i}$ is a complete Haussdorff space, immidiately
lead to the conclusion that $\cH$ should be {\it metrizable} due to the
Urysohn lemma (see e.g\cite{Sim}). Illustrative examples of a similar type
metrisability has been given for different evolutional classifications
of the species, for data analysis. But since the results are general,
this can be also applied to fundamental structures of quantum space-time.

\paragraph{Time.} We have already commented on the role of time
The role of time in the above considered model may be two-fold.
{\it First, it may be taken as just an evolution parameter}, related to
some clocks which are either external to the system or, being installed at
some deeper level of the same system, have no effect on the dynamics
of the observed level. To some extent, it is similar to an atomic clock
used to study the dynamics of terrestrial objects. This we will
call the classical, or nonrelativistic time point of view.

The other possibility, we call it {\it quantum time point of view},
is to understand time as a discrete coordinate which counts the
number of branching points, and can be properly defined only between
objects (=events) of the same evolution branch. We call it time-like
distance. 
\begin{center}
\begin{figure}[t]
{\centering \epsfig{file=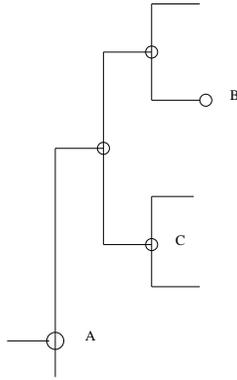, height=5cm} }
\caption{An illustrative binary tree structure of the 
         metric: The distance between two events (objects) is 
         proportional to the age of their common ancestor, 
         i.e. to the time elapsed since they branched off. 
         $T_{AB}=3, T_{BC} = 2$.}
\label{tree:pic}
\end{figure} 
\end{center}
The illustration is
presented in Fig.2. The time distance $T_{AB} = 3$ there. The distance
of this kind can be also defined between two events on different
branches, \ie between causually un-connected events, this we will
call space-like distance. At the Big Bang limit the first ("classical")
approach should be transformed into the second ("quantum") one.

\section{Inertial mass in quantum mechanics defined on a discrete
set}
In standard quantum mechanics we start with the space of
square-integrable functions $\lr$, the evolution operator
$U(t)$, such that
$ \hat U(t) \hat U(t') = \hat U(t+t'),$
and also some representation of canonical comutational relations.
The latter, imply the existence of canonical momentum operator
$\hat p =-\imath\hbar \d /\d x$ defined on a locally Euclidian
(or Minkovskian) coordinate space. The coordinates are then used
to construct a basis in the Hilbert space of state vectors,
viz. $\psi(x) = \bra \psi | x \ket$. The possibility of all these
constructions is not self-evident for arbitrary discrete sets, but
the basic principles are still applicable.
Using the $p$-adic toy model,
we are now going to show hereafter, how the concept of the mass can
be derived in quantum mechanics of a discrete set.

Let $|i\ket$ be a complete set of base observables, $|\psi(t)\ket \in \cH$
be a time-evolving state vector, then
$$ C_i(t) = \bra i |\psi(t) \ket \in \lr$$ is a state function.
If a $p$-adic system is considered the basic vectors in $\cH$
can be labeled by $p$-adic numbers $i\in\Qp$ and $\bra i |\psi(t) \ket$
is the amplitude of the probability to find {\it an excitation of
the i-th object}. The evolution operator can be written in usual form:
\be
U_{ij}(t+\Delta t,t) = \delta_{ij} - \frac{1}{\hbar} \Delta t H_{ij},
\label{ev}
\ee
where $i,j\in\Qp$ and time is considered classically. Taking the scalar
product of $\bra i|$ and the action of the operator (\ref{ev}) to the
vector $|j\ket$ one gets
$$ \imath\hbar \frac{dC_i(t)}{dt} = \sum_j H_{ij}(t) C_j(t),\quad i,j\in\Qp, $$
or, using the convenient notation of $p$-adic integration
\be
\imath\hbar \frac{dC(i)}{dt} = \int_\Qp H(i-j,t) C(j) dj.
\label{pse}
\ee
Here we to make some physical asumptions about the
Hamiltonian in (\ref{pse}). First, in the spirit of our geometrical
interpretation, we assume that quantum transitions ocure between
equidistant objects ($|x|_p=|y|_p$),\ie they correspond to $p$-adic
free particle motions (translations):
$ x'=x+a,\quad\hbox{where }|x|_p=|a|_p.$
Thus, in the simplest case we consider only the transitions within
fixed $p$-adic circles. The system (\ref{pse}) can be then rewritten in the
form
\be
\imath\hbar \frac{dC_i}{dt} = \sum_{j=0}^{p-1} H_{ij}(t) C_j(t), \quad
0 \le i,j < p
\ee
($p>2$ to get a physically sensible theory).
Further, taking $p=3$ to get the most simple model, we follow
the consideration \cite{BGS94}, and after some simple algebra arrive
at the equaion
\be
\imath\hbar \frac{dC(n)}{dt} = E C(n) - A C(n+1) - A C(n-1),\quad
n\in\Q_3,
\label{B}
\ee
where the grownd state of the system can be chosen such, that $E=2A$
\cite{Feyn}.

Now, instead of a Taylor expansion, as was done in \cite{BGS94},
we take a $p$-adic Fourier transform of the \rhs of (\ref{B}).
For convenience, we rewrite it in the form \rhs(\ref{B}) =
$-A(C(n+1) - C(n)+C(n-1)-C(n))$
\begin{eqnarray*}
 C(n\pm1) - C(n) &=& \int_\Qp \bigl[
\chi_p(k(n\pm1)) - \chi_p(kn) \bigr] \tilde C(k) dk \\
&=& \int_\Qp \chi_p(kn)\bigl[\chi_p(\pm k)-1\bigr] \tilde C(k) dk
\end{eqnarray*}
So,
\begin{eqnarray}
\nonumber C(n+1)-2C(n)+C(n-1) &=& 2\int_\Qp \chi_p(kn)\Bigr[
 \frac{\chi_p(-k)+\chi_p(k)}{2}-1 \Bigl] \tilde C(k) dk \label{cd} \\
    &=& 2\int_\Qp \chi_p(kn)\bigl[\cos k -1\bigr] \tilde C(k) dk
\end{eqnarray}
As $n\in \Zp$ is taken, the latter equation is non zero only for
$\{k\}_p \ne 0$, \ie $k\in \Zp^{-1}$, hence $k$ and $n$ are dual
to each other in the sense that $\Zp\oplus\Zp^{-1}=\Qp.$
The equality $\frac{\chi_p(-k)+\chi_p(k)}{2} = \cos k$, always
holds for the multiplication $k \cdot -1$
affects only the {\it integer} part of $k$ and the factor $2\pi\times
integer$ can be droped in the exponent.
Expanding the cosine into the Taylor series
$$\cos k = \sum_{l=0}^\infty \frac{(-1)^l}{(2l)!}k^{2l}$$
and keeping the terms up to the first order in $l$ we obtain
a familiar form of the Schr\"odinger equation for a free
particle
\be
\imath\hbar \frac{dC(n)}{dt} = -A\int \chi_p(kn) k^2\tilde C(k) dk.
\label{se1}
\ee

It is important to emphasize here, that having started with a discrete
system in (\ref{B}) we arrive at the same Schr\"odinger equation as
in continuous theory, but without any asumptions as to the existence
of continuous space. If $n\in\Zp$ labels the constituentsof the system,
as was implied above, then $k\in\Zp^{-1}$ can be understood as $p$-adic
momentum. As in standard quantum mechanics, the
energy of a free excitation  here is $\frac{\hbar^2k^2}{2m'}$, but
it must be also equal to the energy term in the \rhs of the eq.(\ref{se1}):
$$\frac{\hbar^2k^2}{2m'} = Ak^2.$$
Hence we can express the mass of the excitation in the quantum system
(\ref{B}) in terms of the probability amplitude $A$, and a dimensional
constant $\hbar$:
\be
m' = \frac{\hbar^2}{2A}.
\label{mass}
\ee
The time $t$ in eq.(\ref{se1}) is understood as an evolution
parameter only, and we can choice $\hbar=1$ system of units.
{\it The mass of the excitation is therefore, up to a dimensional
constant, just an iverse of transition probability amplitude}.

\section{Conclusion}
The advantage of the above approach is that it is {\it coordinate-free}
and thus no asumptions about the existence of any space before or beyond
matter were used. We have shown that if there exists a set
of objects which may be labeled by $p$-adic numbers --- the asumption is
fairly general --- we can introduce the mass as the inverse probability
amplitude of transition between these objects, and further, the concept of
metric also naturally arises.
%%%%%%%%%%%%%%%%%%%%%%%%%%%%%%%%%%%%%%%%%%%%%%%%%%%
\def\cmp{{\it Comm. Math. Phys.}}
\def\cqg{{\it Class. Quant. Grav.}}
\def\mpla{{\it Mod. Phys. Lett.}A}
\def\ijtp{{\it Int. J. Theor. Phys.}}
\def\prl{{\it  Phys. Rev. Lett.}}
\def\prd{{\it  Phys. Rev.}D}
\def\npb{{\it  Nucl. Phys.}B}

\end{document}